\documentclass{article}

\usepackage{mathptmx, graphics, graphicx, verbatim, url, amssymb, amsmath, amsfonts, amsthm, latexsym}

\newtheorem{theorem}{Theorem}

\begin{document}

\title{The temporalized Massey's method}

\author{Enrico Bozzo \\
Department Mathematics, Computer Science, and Physics \\ 
University of Udine \\
\url{enrico.bozzo@uniud.it} \and 
Massimo Franceschet\\
Department Mathematics, Computer Science, and Physics \\ 
University of Udine \\
\url{massimo.franceschet@uniud.it}}

\maketitle

\begin{abstract}
We propose and throughly investigate a temporalized version of the popular Massey's technique for rating actors in sport competitions. The method can be described as a dynamic temporal process in which team ratings are updated at every match according to their performance during the match and the strength of the opponent team. Using the Italian soccer dataset, we empirically show that the method has a good foresight prediction accuracy.
\end{abstract}

\section{Introduction} \label{sec:introduction}

Rating and ranking in sport have a flourishing tradition. Each sport competition has its own official rating, from which a ranking of players and teams can be compiled.  The challenge of many sports' fans and bettors is to beat the official rating method: to develop an alternative rating algorithm that is better than the official one in the task of predicting future results. As a consequence, many sport rating methods have been developed. Amy N. Langville and Carl D. Meyer even wrote a (compelling) book about (general) rating and ranking methods entitled \textit{Who's \#1?} \cite{LM12}.

In 1997, Kenneth Massey, then an undergraduate, created a method for ranking college football teams. He wrote about this method, which uses the mathematical theory of least squares, as his honors thesis \cite{M97}. Informally, at any given time $t$, Massey's method rates a team $i$ according to the following two factors: (a) the difference between points for and points against $i$, or point spread of $i$, up to time $t$, and (b) the ratings of the teams that $i$ matched up to time $t$. Hence, highly rated teams have a large point differential and matched strong teams so far. Below in the ranking are teams that did well but had an easy schedule as well as teams that did not so well but had a tough schedule.

In this paper we propose a temporalized version of the original Massey's method. The idea is the following. For a given team $i$ and time $t$, the original Massey rates $i$ according to the point spread of $i$ up to time $t$ and the ratings of the teams that $i$ matched up to time $t$. Notice, however, that the rating of a matched team $j$ is computed with respect to time $t$, and not, as we argue it should be more reasonable, with respect to the (possibly previous) time when $i$ and $j$ matched. Suppose, for instance, that $i$ and $j$ matched at time $7$, when team $j$ was strong (high in the ranking), and now, at time $19$, team $j$ lost positions in the ranking and is thus weaker. The original Massey's method adds up to the rating of $i$ the \textit{current} low rating of $j$ computed at time $19$, and not the \textit{past} high rating of $j$ computed at time $7$. The temporalized Massey's method we propose solves this issue. At any given time $t$ of the season, the temporalized Massey's method rates a team $i$ according to (a) the point spread of $i$ up to time $t$, and (b) the ratings of the teams that $i$ matched up to time $t$ computed with respect to the \textit{time they matched}.

The paper is organized as follows. Section \ref{sec:massey} reviews the original Massey's method. We propose the temporalized interpretation of the Massey's method in Section \ref{sec:tmassey}. In Section \ref{sec:closer} we investigate the algebra of the proposed method while in Section \ref{sec:example} we apply it to the last Italian soccer championship.  We review related methods for sport rating in Section \ref{sec:related}. Finally, we conclude in Section \ref{sec:discussion}.

\section{The Massey's method for sports ranking} \label{sec:massey}
In this section we offer a brief introduction to the original Massey's method.
A more general introduction can be found in \cite{GS16}.
The main idea of Massey's method, as proposed in \cite{M97}, is enclosed in the following equation:

$$r_i - r_j = y_k$$
where $r_i$ and $r_j$ are the ratings of teams $i$ and $j$ and $y_k$ is the margin of victory for game $k$ of team $i$. If there are $n$ teams who played $m$ games, we have a linear system:

\begin{equation} \label{Massey1}
X r = y
\end{equation}
where $X$ is a $m \times n$ matrix such the k-th row of $X$ contains all 0s with the exception of a 1 in location $i$ and a $-1$ in location $j$, meaning that team $i$ beat team $j$ in match $k$ (if match $k$ ends with a draw, either $i$ or $j$ location can be assigned $1$, and the other $-1$). Observe that, if $e$ denotes the vector of all $1$'s, then $Xe=0$.
Let $M = X^T X$ and $p = X^T y$. Notice that

\begin{equation*}
M_{i,j} = \left\{
    \begin{array}{ll}
     \text{the negation of the \# of matches between } i \text{ and } j & \text{if } i \neq j,\\
      \text{\# of games played by } i  & \text{if } i = j.
    \end{array} \right.
\end{equation*}
and $p_i$ is the signed sum of point spreads of every game played by $i$. Clearly the entries of $p$ sum to $0$, in fact $e^Tp=e^TX^Ty=(Xe)^Ty=0$. The Massey's method is then defined by the following linear system:

\begin{equation} \label{Massey2}
M r = p
\end{equation}
which corresponds to the least squares solution of system (\ref{Massey1}).

We observe how the Massey's team ratings are in fact interdependent. Indeed, Massey's matrix $M$ can be decomposed as

$$M = D - A,$$
where $D$ is a diagonal matrix with $D_{i,i}$ equal to the number of games played by team $i$, and $A$ is a matrix with $A_{i,j}$ equal to the number of matches played by team $i$ against team $j$. Hence, linear system (\ref{Massey2}) is equivalent to

\begin{equation} \label{Massey3}
Dr - Ar = p,
\end{equation}
or, equivalently

$$
r = D^{-1} (A r + p) = D^{-1} A r + D^{-1} p.
$$
That is, for any team $i$

\begin{equation} \label{Massey4}
r_i = \frac{1}{D_{i,i}} \sum_j A_{i,j} r_j + \frac{p_i}{D_{i,i}}.
\end{equation}
This means, and the same observation can be found in \cite{GS16}, that the rating $r_i$ of team $i$ is the sum $r^{(1)}_{i} + r^{(2)}_{i}$ of two meaningful components:

\begin{enumerate}
\item the mean rating of teams that $i$ has matched $$r^{(1)}_{i} = \frac{1}{D_{i,i}} \sum_j A_{i,j} r_j;$$
\item the mean point spread of team $i$ $$r^{(2)}_{i} = \frac{p_i}{D_{i,i}}.$$

\end{enumerate}
It is worth pointing out that the ratings computed by Massey's method correspond to averages. Hence, it could happen that a team that plays with good performances a limited number of matches against strong teams obtains an extremely high and not justified rating. Actually this effect has been clearly discussed in \cite{CHH14}.
To overcome this problem the authors propose to introduce a dummy team that defeats all the teams that played a number of matches below a suitable cutoff.

In order to better understand the behaviour of the method,
it is interesting to analyse what happens to Massey's system at the end of the season, assuming a round-robin competition in which all $n$ teams matched all other teams exactly once. In this case, the opponents rating component $$r^{(1)}_{i} = -\frac{r_i}{n-1},$$ where we have used the fact that $\sum_i r_i = 0$, and the point spread component $$r^{(2)}_{i} = \frac{p_i}{n-1},$$ hence $$r_i = r^{(1)}_{i} + r^{(2)}_{i} = -\frac{r_i}{n-1} + \frac{p_i}{n-1},$$ and thus $$r_i = \frac{p_i}{n}.$$ Hence, the final rating of a team is simply the mean point spread of the team.
It is possible to be a bit more precise about this property of Massey's method by exploiting the properties of the set of eigenvalues, or spectrum, of the Laplacian matrix $M = D - A$. The spectrum
reflects various aspects of the structure of the graph $G_A$ associated with $A$, in particular those related to connectedness.  It is well known that the Laplacian is singular and positive semidefinite (recall that $M=X^TX$ and $Xe=0$)
so that its eigenvalues are nonnegative and can be ordered as follows:
$$\lambda_1=0\le \lambda_2\le \lambda_3 \le \ldots\le \lambda_n.$$
It can be shown that $\lambda_n\le n$, see for example \cite{BH12}.
The multiplicity of $\lambda_1=0$ as an eigenvalue of the Laplacian can be shown to be equal
to the number of the connected components of the graph, see again  \cite{BH12}.
If the graph of the matches is connected or, equivalently,  $M$ is irreducible, as we assume in the following, $\lambda_2\neq 0$ is known as {\em algebraic connectivity} of the graph and is an indicator of the effort to be employed in order to disconnect the graph.

We can write the spectral decomposition of $M$ as $M=UDU^T$ where $U$ is orthogonal and its first column is equal to $e/\sqrt{n}$, and $D={\rm diag}(0$, $\lambda_2$, $\ldots$, $\lambda_n)$. From $Mr=p$ we obtain $r=UD^+U^Tp$ where
$D^+={\rm diag}(0$, $\frac{1}{\lambda_2}$, $\ldots$, $\frac{1}{\lambda_n})$.
Now
$$r-\frac{p}{n}=UD^+U^Tp-\frac{p}{n}=U\Bigr[D^+-\frac{I}{n}\Bigl]U^Tp,$$
where $I$ is the identity matrix.
Observe that the first component of the vector $U^Tp$ is equal to zero so that
$$r-\frac{p}{n}=U\Bigr[D^+-\frac{I}{n}\Bigl]U^Tp=U\Bigr[D^+-\frac{\tilde{I}}{n}\Bigl]U^Tp,$$
where $\tilde{I}={\rm diag}(0,1,\ldots,1)$. If we denote with $\|\cdot\|$ the Euclidean norm we obtain
$$\|r-\frac{p}{n}\|=\|U\bigr[D^+-\frac{\tilde{I}}{n}\bigl]U^Tp\|\le \|p\| \max_{k=2,\ldots,n} \Bigl| \frac{1}{\lambda_k}-\frac{1}{n}\Bigr|\le \|p\|\frac{n-\lambda_2}{n\lambda_2},$$
where we used the fact that the Euclidean norm of an orthogonal matrix is equal to one.
Hence, as the algebraic connectivity $\lambda_2$, as well as the other eigenvalues, approach $n$, that is, as more and more matches are played,
the vector $r$ approaches $p/n$ and the equality is reached when the graph of the matches becomes complete.

\section{Temporalized Massey's method} \label{sec:tmassey}
We propose a temporalized variant of the original Massey's method. The main idea of the new proposal is to compute the rating of a matched team with respect to the time when the match was played, and not with respect to the current time, as Massey does.

We consider a temporal process of matches between pairs of teams that occur at a given time. Each element of the process is a tern $(i,j,t)$ where $i$ and $j$ are the teams that matched and $t$ is the time of the match. Time is discrete and is represented with natural numbers $0, 1, \ldots$. We assume that each team plays at most one match at any given time. Matches (of different teams) that occur at the same time are considered to happen simultaneously.

\begin{comment}
Let $p_i(t)$ be the cumulative point spread of team $i$ at time $t$. We have $p_i(0) = 0$ for all teams $i$ and  for $t \geq 1$:

$$p_i(t) = p_i(t-1) + s_i(t)$$
where $s_i(t)$ is the difference of the points for team $i$
and the points against team $i$ in the match of time $t$ ($s_i(t)$ is assumed equal to 0 if $i$ did not play at time $t$).
\end{comment}
Let $s_i(t)$ be the difference of the points for team $i$  and the points against team $i$ in the match of time $t$, where we assume $s_i(t) = 0$ if $i$ does not play at time $t$. Let $m_{i,t}$ be the number of games that team $i$ played until time $t$. Let $j_1, \ldots, j_{m_{i,t}}$ be the teams matched by $i$ until time $t$ and $t_1, \ldots, t_{m_{i,t}}$ be the timestamps of these matches. Then the rating of team $i$ at time $t$ is defined as follows. We set $r_i(0) = 0$ for all teams $i$. Hence all teams are initially equally ranked. For any team $i$, if $i$ did not play so far, that is $m_{i,t} = 0$, then its rating is still null. Otherwise, if $m_{i,t} > 0$, we have that, for every $t \geq 1$:

\begin{equation} \label{eq:tmassey0}
r_i(t) = \frac{1}{m_{i,t}} \sum_{k=1}^{m_{i,t}} (r_{j_k}(t_k - 1)+s_i(t_k)).
\end{equation}
This means that the rating $r_i(t)$ of team $i$ at time $t$ is the sum $r^{(1)}_{i}(t) + r^{(2)}_{i}(t)$ of two meaningful components:

\begin{itemize}
\item the mean \textit{historical} rating of teams that $i$ has matched: $$r^{(1)}_{i}(t) = \frac{1}{m_{i,t}} \sum_{k=1}^{m_{i,t}} r_{j_k}(t_k - 1);$$

\item the mean point spread of team $i$ at time $t$: $$r^{(2)}_{i}(t) =\frac{1}{m_{i,t}} \sum_{k=1}^{m_{i,t}} s_i(t_k).$$
\end{itemize}
Notice that we set $r_i(0) = 0$ for all teams, meaning that at the start of the competition all teams are considered equal.
This might be not always realistic: we sometimes know that some teams are potentially stronger than others. Hence, an alternative solution is to set $r_i(0) = \rho_i$, where $\rho_i$ is the exogenous strength of $i$ before the competition starts. For instance, we can set the exogenous strength to be proportional to the rating of the team at the end of the previous season.

We illustrate the proposed method with the following simple example (a complete application is discussed in Section \ref{sec:example}). The table below shows the results of 6 matches (numbered from 1 to 6), divided in 3 days representing a different time (numbered from 1 to 3), involving 4 fictitious teams (labelled A, B, C, D):

\bigskip
\begin{center}
\begin{tabular}{cccccc}
match & day & team 1 & team 2 & score 1 & score 2 \\ \hline
1 & 1 & A  &   C  &   2   &   1 \\
2 & 1 & B  &   D  &   2   &   1 \\
3 & 2 & A  &   D  &   3   &   0 \\
4 & 2 & B  &   C  &   1   &   1 \\
5 & 3 & A  &   B  &   1   &   0 \\
6 & 3 & C  &   D  &   1   &   0 \\
\end{tabular}
\end{center}
While there is no doubt that A is the leader of the ranking (it won all matches) and D is the weakest team (it lost all matches), the challenge between B and C is more controversial: each has won one match, lost another match and drew when they matched together.

The following spread matrix contains the cumulative spread of each team at each day. Initially B has a small advantage over C, which is maintained in the second day, and lost in the last day, when they finish with the same spread. Notice that the spread of the last day corresponds, up to a multiplicative constant, to  the original Massey rating (see Section \ref{sec:massey}). Hence, according to the spread or to orignal Massey's method, there is no difference between B and C at the end of the season.

\bigskip
\begin{center}
\begin{tabular}{l|rrr}
   & 1 & 2 & 3 \\ \hline
A & 1  & 4  & 5 \\
B & 1  & 1  & 0 \\
C & -1 & -1 & 0 \\
D & -1 & -4 & -5 \\
\end{tabular}
\end{center}
However, the temporalized Massey's method tells us a different story. The following matrix contains the temporalized Massey rating for each day and each team:

\bigskip
\begin{center}
\begin{tabular}{l|rrr}
   & 1 & 2 & 3 \\ \hline
A & 1  & 1.5  & 1.33 \\
B & 1  & 0    & 0.17 \\
C & -1 & 0    & -0.17 \\
D & -1 & -1.5 & -1.33 \\
\end{tabular}
\end{center}
The first day the rating is exactly the spread, hence B has an little advantage over C. Interestingly, this advantage is lost at day 2, while the spread is still in favor of B. The reason is that at day 2, teams B and C matched together and they drew. However, before of the match (at day 1), B was stronger than C, hence C drew against a stronger team with respect to B. Finally, at day 3, B is over C in the ranking (while the spread is equal). In fact, at day 3, B lost, but against the strongest team of the competition (A), and C won, but against the weakest team of the competition (D). In summary, B and C drew the match together (but when B was stronger), and then they both lost against A and won against D. But the subtle difference, which is captured only by the temporalized version of Massey, is that B lost against A at day 3, when A was the strongest team, while C lost against A at day 1, when A was as strong as all other teams. Similarly, B won against D at day 1, when D was as strong as all other teams, while C won against D at day 3, when D was the weakest team. This determines the difference in the final ranking of the temporalized Massey's method.

\subsection{A closer look to temporalized Massey's method} \label{sec:closer}

Let us consider more closely the temporalized Massey's equation (\ref{eq:tmassey0}).
Clearly, if at time $t$ team $i$ does not play then $r_i(t)=r_i(t-1)$. On the contrary,
suppose that at time   $t$  team $i$ matches with team $j$ (in other words $t=t_k$ for some $k$). Then the rating of $i$ at time $t$ can be defined in terms of the ratings at $t-1$ of teams $i$ and $j$ as well as the point spread of team $i$ at the current time $t$:

\begin{equation} \label{eq:tmassey1}
r_i(t) =  \frac{m_{i,t}-1}{m_{i,t}} r_i(t-1) + \frac{s_i(t) + r_{j}(t-1)}{m_{i,t}}.
\end{equation}
Similarly, the rating of $j$ at time $t$ is:

\begin{equation} \label{eq:tmassey1.1}
r_j(t) =  \frac{m_{j,t}-1}{m_{j,t}} r_j(t-1) + \frac{s_j(t) + r_{i}(t-1)}{m_{j,t}}.
\end{equation}
Notice that losing against a strong team can still make the day for the loser, but winning against a weak team can result is a drop of the rating of the winner. We can rewrite Equation \ref{eq:tmassey1} as follows:

\begin{equation} \label{eq:tmassey2}
r_i(t) =  \alpha_{i,t} \, r_i(t-1) + \beta_{i,t} \, r_{j}(t-1) + \beta_{i,t_k} \, s_i(t),
\end{equation}
where $\alpha_{i,t} = (m_{i,t}-1)/m_{i,t}$ and $\beta_{i,t} = 1/m_{i,t}$. Notice that $\alpha_{i,t} + \beta_{i,t} = 1$. Hence, the rating of team $i$ at time $t$ is a convex combination of the ratings at time $t-1$ of teams $i$ and of the matched team $j$ plus a fraction of the spread of $i$ at time $t$. Of course, by expanding recurrence (\ref{eq:tmassey2}) one obtains back equation (\ref{eq:tmassey0}).

We would like to attract the attention of the reader to the fact that coefficients $\alpha_{i,t}$ and $\beta_{i,t}$ vary in time. More precisely, as the number of games $m_{i,t}$ of team $i$ grows, the component $\alpha_{i,t}$ approaches $1$ and $\beta_{i,t}$ vanishes to $0$. This means that, if $i$ played few matches and hence $m_{i,t}$ is small, then the latest performance of $i$ can make a significant difference in the ranking position of team $i$. On the other hand, as $m_{i,t}$ grows, new results can only slightly move the ranking position of the team. This is coherent with the general idea that an established reputation is difficult to shake.

Interestingly, if teams $i$ and $j$ played the same number of matches at time $t$, that is $m_{i,t} = m_{j,t}$, it is easy to realize that, after a match between $i$ and $j$, we have that  $r_i(t) + r_j(t) = r_i(t-1) + r_j(t-1)$. This means that what one team gains is lost by the other, and the cumulative rating of the system is the same before and after the match. In particular, in a round-robin competition in which at each day in the competition each team matches another team not matched before, it happens that, if initially all teams have rating equal to 0, at any day the cumulative rating of all teams in the competition is 0. It is worth noticing that this property holds also for the original Massey's method but is lost if teams play a different number of games.

\begin{comment}
$$
\begin{array}{ll}
r_i(t) + r_j(t)  & =  \alpha_t \, r_i(t-1) + \beta_t \, r_{j}(t-1) + \beta_t \, s_i(t) +
\alpha_t \, r_j(t-1) + \beta_t \, r_{i}(t-1) - \beta_t \, s_i(t) \\
& = (\alpha_t + \beta_t) \, r_i(t-1) + (\alpha_t + \beta_t) \, r_{j}(t-1) \\
& = r_i(t-1) + r_{j}(t-1)
\end{array}
$$
\end{comment}

From (\ref{eq:tmassey1}) it follows that every rating $r_i(t)$ is a linear combination of spreads  whose nonnegative coefficients  can
be placed in a matrix $C^{(i,t)}$ such that
$$r_i(t)=\sum_{k=1}^n\sum_{l=1}^t    C^{(i,t)}_{k,l}s_{k}(l).$$
From (\ref{eq:tmassey1}) it is possible to obtain an equivalent relation for these matrices  in the case where $i$ matches with $j$ at time $t$
\begin{equation} \label{eq:matrices}
C^{(i,t)} =  \frac{m_{i,t}-1}{m_{i,t}} C^{(i,t-1)} +
\frac{1}{m_{i,t}}E^{(i,t)} + \frac{1}{m_{i,t}}C^{(j,t-1)},
\end{equation}
where $E^{(i,t)}_{k,l}=1$ if $(i,t)=(k,l)$ and $E^{(i,t)}_{k,l}=0$ otherwise. Clearly only the first $t$ columns of $C^{(i,t)}$ contain entries different from zero.

As an example let us consider again the $4$ fictitious teams $A$, $B$, $C$ and $D$ of the previous example that now is convenient to denote with the integers from $1$ to $4$. In this simple example every team plays at each time hence $m_{i,t}=t$. Therefore Equation (\ref{eq:matrices})  becomes
\begin{equation} \label{eq:matrices2}
C^{(i,t)} =  \frac{t-1}{t} C^{(i,t-1)} +
\frac{1}{t}E^{(i,t)} + \frac{1}{t}C^{(j,t-1)},\qquad t=1,2,3
\end{equation}
and this yields

\begin{equation}
C^{(1,1)}=\begin{bmatrix}1\\0\\0\\0\end{bmatrix},\quad C^{(1,2)}=\begin{bmatrix}1/2& 1/2\\0 &0 \\0& 0\\1/2& 0\end{bmatrix}, \quad
C^{(1,3)}=\begin{bmatrix}1/3& 1/3 &1/3\\1/6 &1/6 & 0 \\1/6 &0& 0\\1/3& 0 & 0\end{bmatrix}
\end{equation}
where only the nontrivial columns of the matrices are shown. Of course if the $4$ teams are involved in a round robin competition then in the 4th day $A$ and $C$ match together again and
$$
C^{(1,4)}=\begin{bmatrix}7/24& 1/4 &1/4 & 1/4\\5/24 &1/8 & 0 & 0
                      \\5/24 &1/12& 1/12 & 0\\7/24& 1/24 & 0 & 0\end{bmatrix},
$$
where, as before, only the nontrivial columns of the matrix are shown.
It is possible to verify that $C^{(i,t)}$ for $i=2,3,4$ are just row permutations of $C^{(1,t)}$.

Notice that the sum of the coefficients in the columns of the matrices $C^{(i,t)}$ in our example  has a quite regular behaviour.  Let us denote with $C^{(i,t)}_{:,l}$
the $l$-th column of $C^{(i,t)}$. By using (\ref{eq:matrices2}), for $l=t$ we obtain $$e^TC^{(i,t)}_{:,t}=\frac{1}{t}e^TE^{(i,t)}_{:,t}=\frac{1}{t},$$ that is true in particular for $l=t=1$. Making use of induction we obtain for $l\le t-1$
\begin{eqnarray*}
e^TC^{(i,t)}_{:,l}=\frac{t-1}{t}e^TC^{(i,t-1)}_{:,l}+\frac{1}{t}e^TC^{(j,t-1)}_{:,l}                  =\frac{t-1}{t}\frac{1}{l}+\frac{1}{t}\frac{1}{l}=\frac{1}{l}.
\end{eqnarray*}
\begin{comment}
\begin{theorem} \label{teo:columnsums}
Let  $m_{i,t}=t$.
It turns out that for $l=1,\ldots,t$
$$e^TC^{(i,t)}_{:,l}=\frac{1}{l}.$$
\end{theorem}
\begin{proof}
Let us proceed by induction on $t$. For $t=1$ the assertion is true. Let us assume that is
true for $t-1$. Then from recurrence (\ref{eq:matrices}) we obtain for $l\le t-1$
\begin{eqnarray*}
e^TC^{(i,t)}_{:,l}=\frac{t-1}{t}e^TC^{(i,t-1)}_{:,l}+\frac{1}{t}e^TC^{(j,t-1)}_{:,l}                  =\frac{t-1}{t}\frac{1}{l}+\frac{1}{t}\frac{1}{l}=\frac{1}{l}.
\end{eqnarray*}
If $l=t$ we obtain $e^TC^{(i,t)}_{:,t}=\frac{1}{t}e^TE^{(i,t)}_{:,t}=\frac{1}{t}$.
\end{proof}
\end{comment}
As a consequence, the sum of the entries of $C^{(i,t)}$ is equal to
$H_t=\sum_{l=1}^{t}\frac{1}{l}$ for each team $i$. The number $H_t$ is known as the $t$-th {\em harmonic number}.
It holds that
$$H_t \min_{\substack{1 \leq k \leq n\\ 1 \leq l \leq t}} s_k(l) \le r_i(t) \le H_t \max_{\substack{1 \leq k \leq n\\ 1 \leq l \leq t}} s_k(l).$$
It is well known that $lim_{t\rightarrow \infty} H_t-\ln t=\gamma$ where $\gamma \approx 0.577$ is known as Euler-Mascheroni constant. This implies that
the range of the ratings of temporalized Massey's method increase very slowly in $t$. For example  $H_{38}\approx 4.2$. Moreover, the above inequality tells us that ratings  and spreads, which are added up in the temporalized Massey's equation, are of the same order of magnitude.

It is worth noticing that the temporalized Massey's rating of team $i$ at time $t$ is a linear combination of past spreads (performances) of all teams, not just of team $i$, with multiplicative coefficients described by matrix $C^{(i,t)}$. This contrasts with the original Massey's rating for team $i$. Indeed, as shown in Section \ref{sec:massey}, as time goes on, the original Massey's rating for $i$ approaches a linear combination of past performances of $i$, without considering the performances of other teams.

It is interesting to observe that, if the teams have exogenous initial strengths, then the linear combination of spreads has to be complemented with a linear combination of them. For example, in order  to compute $r_1(4)$, one has to add to the combination of spreads whose coefficient appear in $C^{(1,4)}$, the value obtained from
$$\frac{7}{24}r_3(0)+\frac{5}{24}r_4(0)+\frac{5}{24}r_1(0)+\frac{7}{24}r_2(0),$$
since the first match of $A$ is against $C$ and the first match of $B$ is against $D$.

Finally, it is useful to compare recurrence (\ref{eq:tmassey2}) with its constant coefficient equivalent, namely:
\begin{equation} \label{eq:tmassey3}
r_i(t) =  \alpha \, r_i(t-1) + \beta \, r_{j}(t-1) + \beta \, s_i(t),
\end{equation}
where now $\alpha,\beta>0$ are constant with $\alpha+\beta=1$, and again $t$ is the timestamp of the match of $i$ with $j$. By expanding this recurrence we obtain
\begin{comment}
$$
r_i(t)=\alpha^{m_{i,t}} r_i(0)+\beta\sum_{k=0}^{m_{i,t}-1} \alpha^k\Bigl( r_{j_{m_{i,t}-k}}(t_{m_{i,t}-k}-1)+s_i(t_{m_{i,t}-k})\Bigr),
$$
\end{comment}
\begin{equation} \label{eq:tmassey4}
r_i(t)=\alpha^{m_{i,t}} r_i(0) +
             \beta \sum_{k=1}^{m_{i,t}} \alpha^{m_{i,t} - k}\Bigl(
             r_{j_k}(t_k - 1) + s_i(t_k)\Bigr),
\end{equation}
where $m_{i,t}$ is the number of games that team $i$ played until time $t$, while $j_1, \ldots, j_{m_{i,t}}$ are the teams matched by $i$ until time $t$, and $t_1, \ldots, t_{m_{i,t}}$ are the timestamps of these matches. Comparing Equations \ref{eq:tmassey0} and \ref{eq:tmassey4}, we capture the difference between the varying and constant coefficient recurrences. In Equations \ref{eq:tmassey0}, past performances of a team are treated homogeneously, while with Equations \ref{eq:tmassey4} the past is progressively forgotten, giving more importance to recent performances, and this forgetfulness is quicker if $\alpha$ is small (close to 0).

To obtain an alternative intuition of this difference we study the matrices $C^{(i,t)}$ for our simple round robin example. It is not difficult to obtain
$$
C^{(1,1)}=\beta\begin{bmatrix}1\\0\\0\\0\end{bmatrix},\quad C^{(1,2)}=\beta\begin{bmatrix}\alpha  & 1\\0 &0 \\0& 0\\ \beta& 0\end{bmatrix}, \quad
C^{(1,3)}=\beta \begin{bmatrix}\alpha^2 & \alpha  & 1 \\ \alpha\beta  &\beta & 0 \\ \beta^2 &0& 0\\ \alpha\beta & 0 & 0\end{bmatrix},
$$
where only the nontrivial columns of the matrices are shown. In addition
$$
C^{(1,4)}=\beta \begin{bmatrix}
\alpha^3 +\beta^3& \alpha^2  &\alpha  & 1\\
\alpha^2\beta +\alpha\beta^2 &\alpha\beta  & 0 & 0\\
\alpha\beta^2+\alpha^2\beta  & \alpha\beta & \beta  & 0\\
\alpha^2\beta +\alpha\beta^2& \beta^2 & 0 & 0\end{bmatrix},
$$
where again only the nontrivial columns  are shown.
Notice that, not taking into account the factor $\beta$, the entries of each column
of these matrices sum up to a power of the binomial $\alpha+\beta$. Since we assumed $\alpha+\beta=1$, we have that, for $l=1,\ldots,t$, $$e^TC^{(i,t)}_{:,l}=\beta.$$
This result highlights the difference between the varying-coefficient and the constant-coefficient techniques: the latter gives progressively more and more importance to the recent matches with respect to the former.

Again, if exogenous initial strengths are present then the linear combination of spreads has to be complemented with a combination of initial strengths. For example in order to compute $r_1(4)$ to the combination of spreads one has to add
\begin{align*}
\alpha^4r_1(0)+&(\alpha^3\beta+\beta^4)r_3(0)+(\alpha^2\beta^2+\alpha\beta^3)r_4(0)\\
+&(\alpha\beta^3+\alpha^2\beta^2)r_1(0)+(\alpha^2\beta^2+\alpha\beta^3)r_2(0).
\end{align*}

\subsection{Application to Italian soccer league} \label{sec:example}

As a more realistic example, we analyse the Italian Serie A soccer league of season 2015-2016. It is a round-robin competition with 20 teams and 38 days (each pair of teams matches twice). 

\begin{figure}[t]
\begin{center}
\includegraphics[scale=0.40, angle=0]{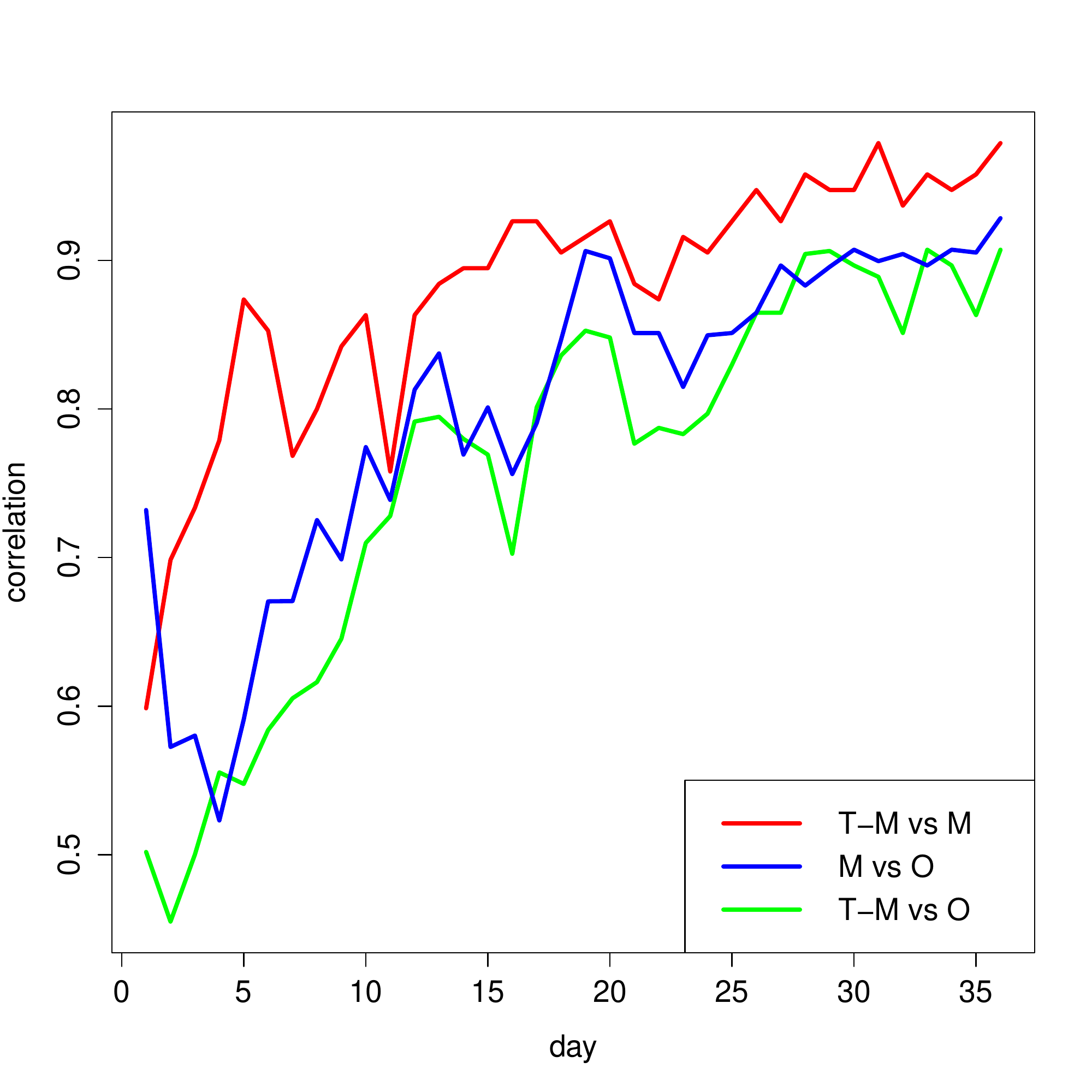}
\end{center}
\caption{The Kendall correlation coefficients among temporalized Massey (T-M), original Massey (M), and the official ranking (O) as days go by from 3 to 38.}
\label{fig:cor}
\end{figure}

In Figure \ref{fig:cor} we depict the Kendall correlation between pairs of ranking methods among temporalized Massey (T-M), original Massey (M), and official ranking (O).
As days pass, we accrue more and more information about the real strength of teams, and all correlations increase. In particular at day 38, end of the season, we have complete information, and correlations coefficients are close to 1 (0.98 for T-M vs M, 0.93 for M vs O, and 0.91 for T-M vs O), although there are differences in the rankings, in particular when the official compilation is involved. Nevertheless, during the season, when information is partial, the corresponding rankings diverge significantly, and correlation coefficients are far from 1, in particular with respect to the official ranking. For instance the coefficients at day 10 are: 0.80 for T-M vs M, 0.73 for M vs O, and 0.62 for T-M vs O. Moreover, over all days, the association between Massey and official rankings is higher than the association between temporalized Massey and official rankings.

\begin{table}[t]
\begin{center}
\begin{tabular}{lll}
\textbf{Method} & \textbf{Without HFA} & \textbf{With HFA} \\ \hline
Temporalized Massey & 0.611  & 0.702 \\ \hline
Elo                 & 0.611  & 0.695 \\ \hline
Official            & 0.589  & 0.674 \\ \hline
\end{tabular}
\end{center}
\caption{Foresight prediction accuracies with and without home-field advantage (HFA). \label{tab:accuracy}}
\end{table}

\begin{figure}[t]
\begin{center}
\includegraphics[scale=0.4, angle=0]{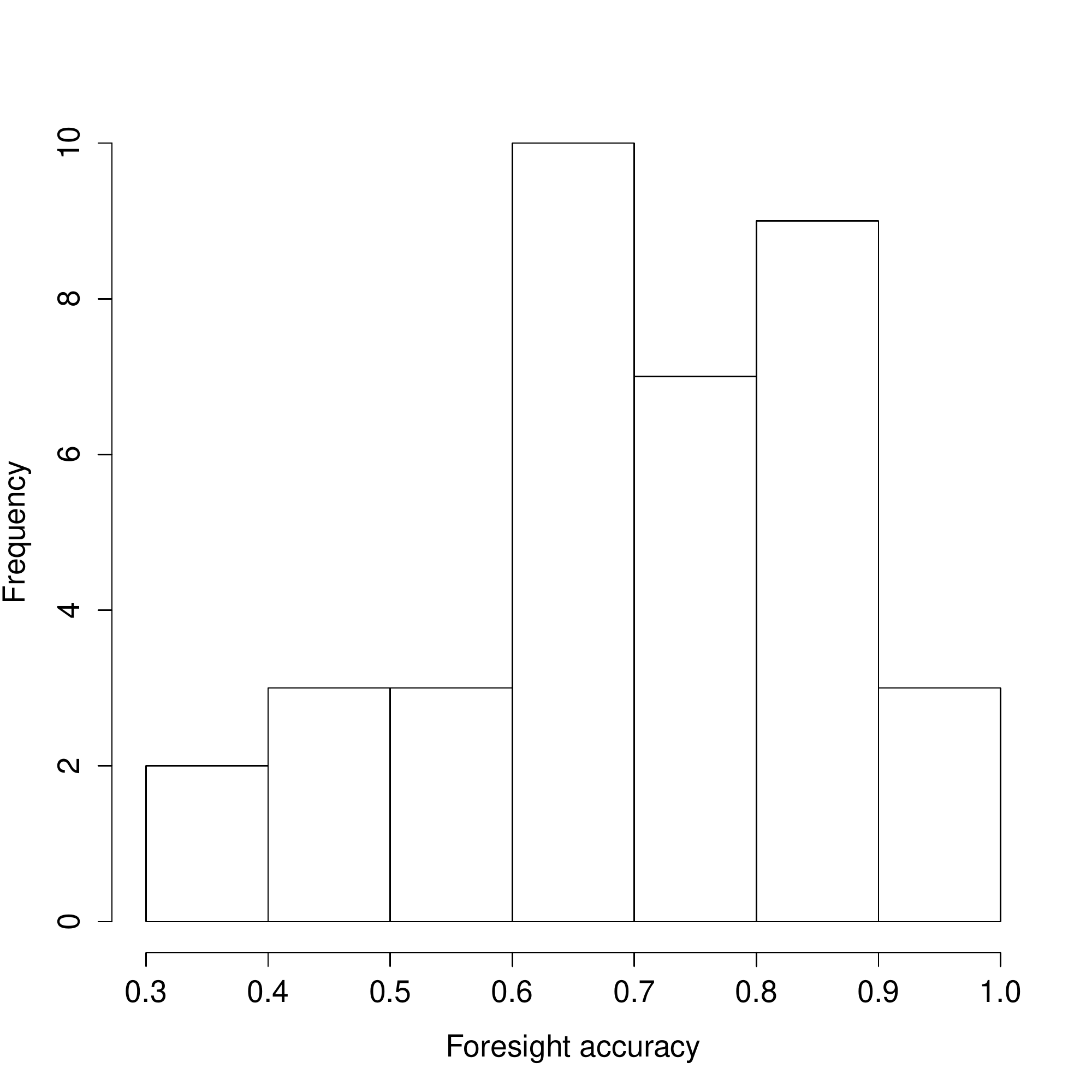}
\end{center}
\caption{Histogram of foresight prediction accuracies at each day of the competition (with home-field advantage) for temporalized Massey's method.}
\label{fig:foresight}
\end{figure}

\begin{figure}[t]
\begin{center}
\includegraphics[scale=0.25, angle=0]{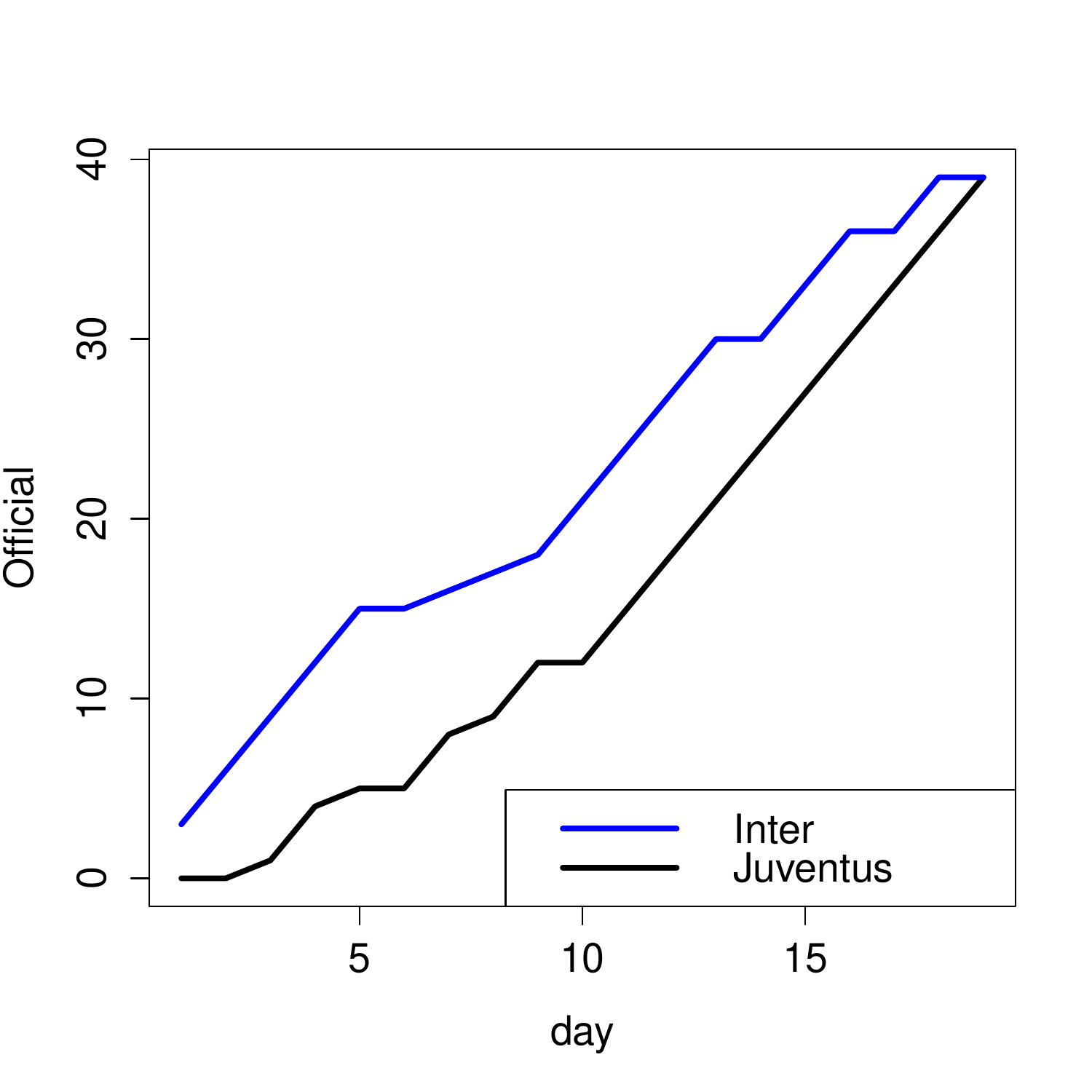}
\includegraphics[scale=0.25, angle=0]{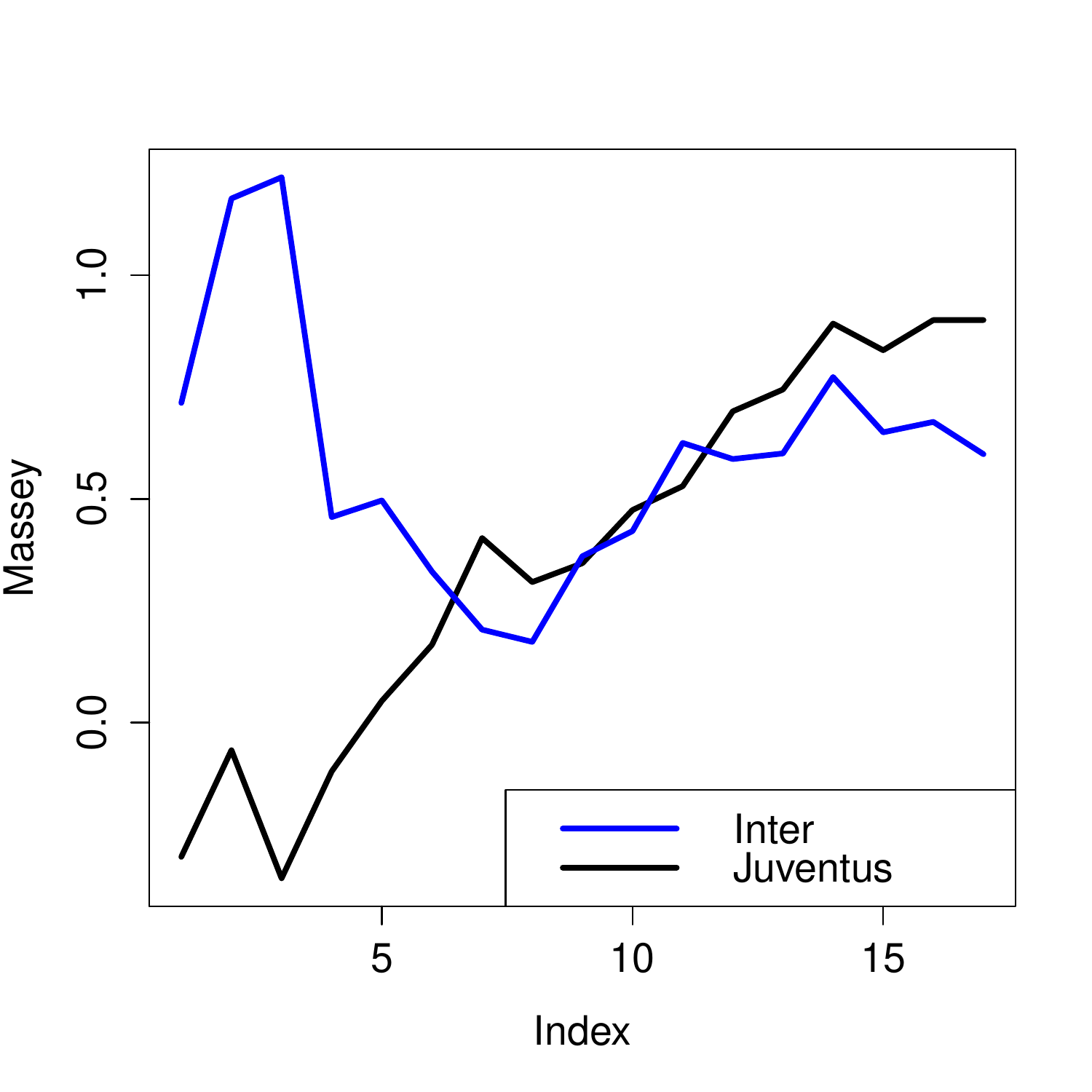}
\includegraphics[scale=0.25, angle=0]{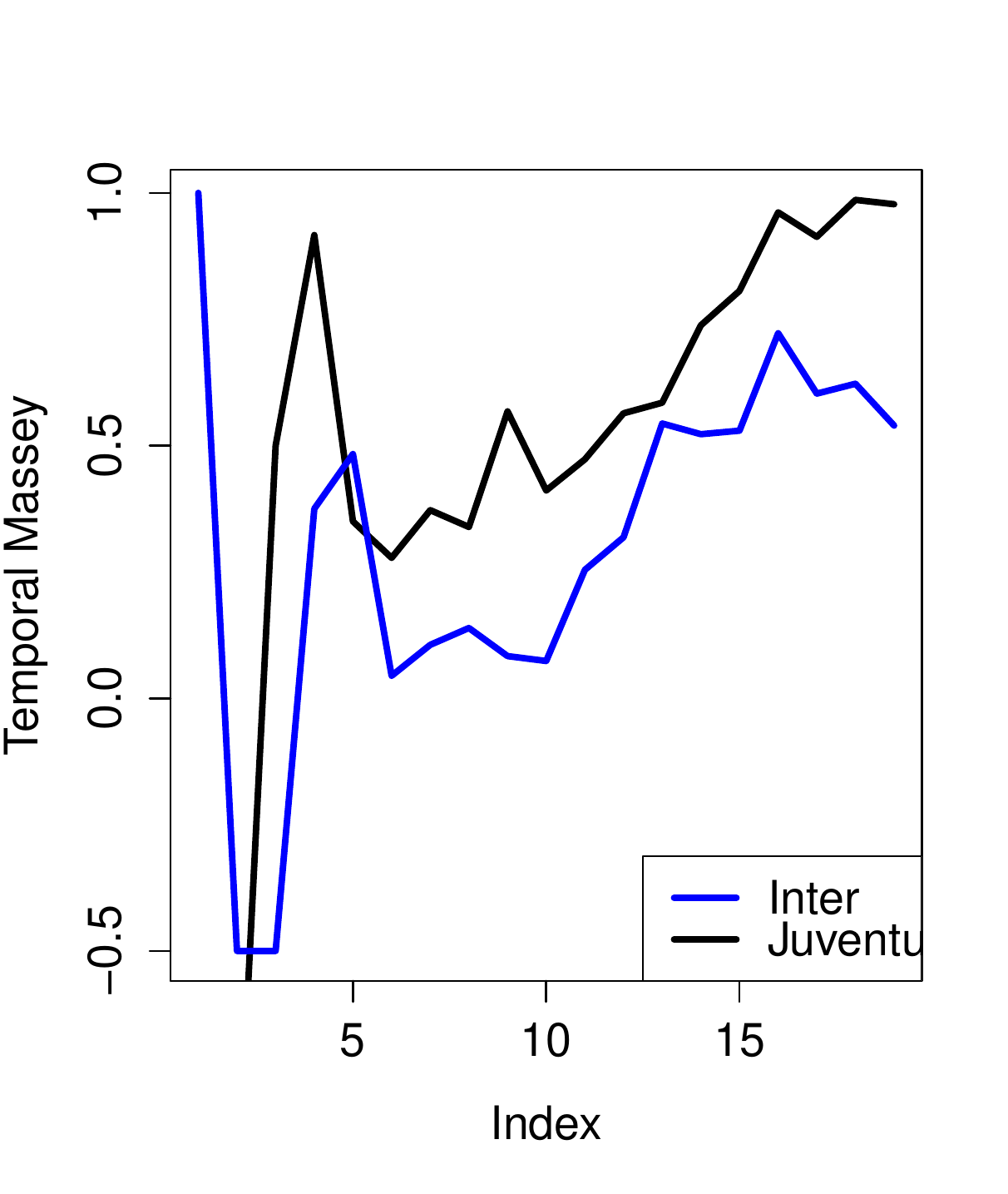}
\end{center}
\caption{The temporal dynamics of the ratings of Juventus and Inter in the first round (19 days).}
\label{fig:IJ}
\end{figure}

A rigorous test for a rating system is \textit{foresight prediction accuracy} \cite{LM12}: how well the vector $r(t)$ of ratings computed at day $t$ can predict the winners at day $t+1$? More precisely, the foresight prediction accuracy of a method is the number of victories that the method corrected foresaw divided by the total number of victories of that competition (we ruled out the ties). Hence, accuracy of 0 means no predictions were correct, while accuracy of 1 means that all predictions were correct. We also computed accuracy introducing a home-field advantage, which was empirically determined for each method and added to the rating of the team playing at home. A home-field advantage matters for foresight prediction in time-varying methods: since initially all teams are rated equal, then in the beginning, before there is enough competition to significantly distinguish the teams' ratings, home-field consideration is the only criterion that the method can use to draw a distinction between two teams. We compared three time-varying rating methods with and without home-field advantage (see Table \ref{tab:accuracy}): official rating of the Italian soccer league, temporalized Massey's method, and Elo's method (see Section \ref{sec:related} for a review of this method). Temporalized Massey is slightly more predictive than Elo and significantly better than the official rating. Moreover, for all methods, introducing the home-field advantage has a significant impact in the prediction accuracy. We also computed, for the temporalized Massey's method, the foresight prediction accuracies at each day of the competition (with home-field advantage). The histogram of accuracies is depicted in Figure \ref{fig:foresight}. Only 2 predictions are below the threshold of 50\% of accuracy corresponding to randomness (notice that the 3 predictions in the 40\%-50\% histogram bar are in fact equal to 50\%). On the other hand, most of predictions (78\%) are above 60\% of accuracy, with 12 predictions (32\%) above 80\% of accuracy and 3 predictions (8\%) with 100\% of accuracy.

\begin{comment}
We computed, day after day, starting at day 3, the accuracy of the methods temporalized Massey, original Massey, and official in predicting the winners of the day after: the higher rated team is predicted to win. The accuracy for a given day is the number of wins that the method predicted corrected divided by the number of wins of that day (we ruled out the ties). Hence, accuracy of 0 means no predictions were correct, while accuracy of 1 means that all predictions were correct.
The aggregated accuracies for all 36 days (excluding the first two days) are given in Table \ref{accuracy}. The two Massey's methods are comparable but both are significantly more accurate in prediction than the official rating (although there was one day in which the official method had 100\% of accuracy).

\begin{table}[t]
\begin{center}
\begin{tabular}{lllll}
\textbf{Method} & \textbf{Min} & \textbf{Median} & \textbf{Mean} & \textbf{Max} \\ \hline
Temporal Massey & 0.2222 &   0.7143 &  0.6438 & 0.8750 \\ \hline
Original Massey & 0.0000 &   0.7143 &  0.6478 & 0.8750  \\ \hline
Official Rating & 0.1667 &   0.6667 &  0.6252 &  1.0000 \\ \hline
\end{tabular}
\end{center}
\caption{Aggregated foresight prediction accuracies. \label{accuracy}}
\end{table}
\end{comment}

Related to prediction accuracy, consider the following story. Teams Inter and Juventus had a peculiar season in 2015-2016. Inter immediately won the first matches, but with low spread of points. On the other hand, the start of Juventus was disastrous. This led Inter well above Juventus in the official ranking, with a maximum distance of 10 points at days 5 and 6. From day 10, however, Juventus started an incredible row of wins, culminating at day 19 when the two teams were pair in official standings. Finally, at day 38, Juventus powerfully won the championship with 24 points above Inter. In Figure \ref{fig:IJ} we depict the temporal dynamics of the official, original Massey, and temporalized Massey rankings during the first round of the championship.
The superiority of Juventus with respect to Inter is not witnessed by the official ranking until the end of the round. On the other hand, Massey and in particular its temporalized version predicted this supremacy well before the end of the round.

\section{Related literature} \label{sec:related}

An recent account of dynamic modelling of sports tournaments can be found in \cite{CVF13}. In the paper, only the outcomes (win-draw-loss) of the matches, and not point spreads, are considered. The abilities of the home and visiting teams are assumed to evolve separately  in time following an exponentially weighted moving average process ruled by a constant coefficients linear recurrence. In our approach the two abilities are twisted together and the evolution is described by a variable coefficients recurrence.

A good survey of dynamic models for teams strengths in NFL can be found in \cite{GS16}.
Generally teams' abilities are assumed to evolve through a first order autoregressive process. For example in \cite{AR80} this strategy is used to model  season to season changes of team' abilities while in \cite{GS98}  week to week changes.
As we explained in Section \ref{sec:tmassey}, due to the variability of the coefficients of recurrence  (\ref{eq:tmassey2}) our approach gives, as season proceeds, a greater importance to the history of the results compared with the one given by an autoregressive model.

In \cite{CKLP11} the authors propose nonuniform weighting for sports rankings.  Their technique allows to weight differently late season play but also, for example, home court advantage or high-pressure games. Actually, their target application is using the matches of the Division I NCAA in order to produce brackets for the famous NCAA Men's Division I Basketball Tournament, also known as March Madness.
For Massey's method this idea is implemented placing the weights in a diagonal matrix $W$ and by solving, instead of system (\ref{Massey1}), the system $X^TWXr=X^TWy$. Notice that this is equivalent to the substitution of the two means present in (\ref{Massey4}) with two weighted means, whose weights are the diagonal entries of $W$. The authors discuss and experiment various strategies for choosing the weights:  in the simplest one the weights linearly increase from the first day of the season to the last day.

The authors also apply their weighting technique to another popular ranking method, namely Colley's method \cite{CO02}. It is important to remark
that the temporalization technique that we developed for Massey's method can easily be extended to Colley's method. The Equation (3) in \cite{CKLP11} is at the heart of Colley's method
and can be rewritten with our notations as follows
\begin{equation}
r_i=\frac{1+(w_i-l_i)+\sum_j A_{i,j}r_j}{2+D_{i,i}},
\end{equation}
where $w_i$  and $l_i$, with $D_{i,i}=w_i+l_i$, are respectively the number of wins and of losses of team $i$.
Our temporalized variant of Colley's method is ruled by the following equation
\begin{equation}
r_i(t)=\frac{1+(w_{i,t}-l_{i,t})+\sum_{k=1}^{m_{i,t}} r_{j_k}(t_k - 1)}{2+m_{i,t}}.
\end{equation}
where now $w_{i,t}$  and $l_{i,t}$, with $m_{i,t}=w_{i,t}+l_{i,t}$, are respectively the number of wins and of losses of team $i$ up and including time $t$.

A popular time-varying rating system used is sport competitions is Elo's method \cite{E78,LM12}. It was coined by the physics professor and excellent chess player Arpad Elo.
\begin{comment}
\footnote{According to the movie \textit{The social network} by David Fincher, it appears that the Elo's method formed the basis for rating people on Zuckerberg's Web site Facemash, which was the predecessor of Facebook.}
\end{comment}
Let $S_{i,j}$ be the score of team $i$ against team $j$; for instance, in chess a win is given a score of 1 and a draw a score of 1/2 (and a defeat a score of 0). Let $\mu_{i,j}$ be the number of points that team $i$ is expected to score against team $j$; this is typically computed as a logistic function of the difference of ratings between the players, for instance, $$\mu_{i,j} = \frac{1}{1 + 10^{-d_{i,j} / \zeta}},$$ where $d_{i,j} = r_i(old) - r_j(old)$ and $\zeta$ is a constant (in the chess world $\zeta = 400$). Then, when teams $i$ and $j$ match, the new rank $r_i(new)$ of team $i$ is updated as follows (and similarly for $j$): $$r_i(new) = r_i(old) + \kappa (S_{i,j} - \mu_{i,j}),$$ where $\kappa$ is a constant (for instance, in chess $\kappa = 25$ for new players).
Hence, beating a stronger player has a larger reward than beating a weaker one. Notice the intriguing similarity of Elo's update equation with Equation (\ref{eq:tmassey2}) defining temporalized Massey's method. Both methods update the old rating of a team in terms of the same ingredients: the current performance of the team and the rating of the opponent team. However, the two methods mix these ingredients in different ways, and hence the resulting recipe differs. While Elo uses a logistic (exponential) function to mix performance and opponent rating, Massey linearly combines the two. Moreover, the combination parameters $\kappa$ and $\zeta$ in Elo are constant, while the combination parameters $\alpha_{i,t}$ and $\beta_{i,t}$ of temporalized Massey vary with the team and in time.

\section{Conclusion} \label{sec:discussion}

We introduced a temporalized version of the popular Massey's method for rating actors in sport competitions. The idea of the new method is quite simple: to rate the matched team with respect to the time when the match was played. We showed that the resulting method can be described as a dynamic temporal process in which the rating of any team $i$ is modified when $i$ matches some other team $j$ and the update of the rating is a function of the performance of $i$ during the match with $j$ and of the rating of $j$ before the match. We applied the new method to the Italian soccer league showing a good foresight prediction accuracy.

In fact, the idea of temporalizing the Massey's method we have proposed in this context can be be generalized to any recursive centrality measures on networks. Consider for instance Pagerank centrality \cite{F11-CACM}, which claims that a node is important if it is linked to by other important nodes. For instance, a scholar is relevant if it is cited by relevant scholars, or a Web page is important if it is hyperlinked to by other important Web pages. The original definition of the Pagerank method ignores the time of creation of the link between nodes. However, we argue that it is different if we, as scholars, receive an endorsement from a young and almost unknown author, or from the same author when she won the Turing award. Similarly, there is a difference in receiving a link from a peripheral Web page or from the same page when it became a central hub. We look forward to a temporalized version of Pagerank with an application to sport competitions.

\bibliographystyle{plain}
%\bibliography{bibliography}

\end{document}